\documentclass[10pt,journal,compsoc]{IEEEtran}
\usepackage[sort&compress,numbers]{natbib}

\ifCLASSINFOpdf
   \usepackage[pdftex]{graphicx}
\else
\fi
\usepackage{amsmath}
\usepackage{algorithm}
\usepackage{algpseudocode}

\usepackage{array}
\usepackage{multirow}
\newcolumntype{P}[1]{>{\centering\arraybackslash}p{#1}}
\newcolumntype{L}[1]{>{\arraybackslash}p{#1}}
\newcolumntype{M}[1]{>{\centering\arraybackslash}m{#1}}
 \usepackage{dblfloatfix}
\usepackage{url}

\newtheorem{definition}{Definition}
\newtheorem{assumption}{Assumption}
\newtheorem{lemma}{Lemma}
\newtheorem{theorem}{Theorem}
\newtheorem{conjecture}{Conjecture}

\begin{document}
%
\title{PALE: Partially Asynchronous Agile\\ Leader Election}
%
%
%

\author{Bronislav~Sidik, Rami~Puzis, Polina~Zilberman, 
        and Yuval Elovici
\thanks{R. Puzis at email: puzis@bgu.ac.il}
\thanks{B. Sidik at email: sidik@post.bgu.ac.il}
\thanks{P. Zilberman at email: polinaz@bgu.ac.il}
\thanks{Y. Elovici at email: elovici@bgu.ac.il}
\thanks{Department of Software and Information Systems Engineering, 
        Ben-Gurion University of the Negev, 
        P.O.B. 653 Beer-Sheva, Israel.}
\thanks{\textbf{This work has been submitted to the IEEE for possible publication. Copyright may be transferred without notice, after which this version may no longer be accessible.}}
}

\maketitle

\begin{abstract}
	
Many tasks executed in dynamic distributed systems, such as sensor networks or enterprise environments with bring-your-own-device policy, require central coordination by a leader node. 
In the past it has been proven that distributed leader election in dynamic environments with constant changes and asynchronous communication is not possible. 
Thus, state-of-the-art leader election algorithms are not applicable in asynchronous environments with constant network changes. 
Some algorithms converge only after the network stabilizes (an unrealistic requirement in many dynamic environments). 
Other algorithms reach consensus in the presence of network changes but require a global clock or some level of communication synchronization. 
	
Determining the weakest assumptions, under which leader election is possible, remains an unresolved problem.
In this study we present a leader election algorithm that operates in the presence of changes and under weak (realistic) assumptions regarding message delays and regarding the clock drifts of the distributed nodes.  
The proposed algorithm is self-sufficient, easy to implement and can be extended to support multiple regions, self-stabilization, and wireless ad-hoc networks. 
We prove the algorithm's correctness and provide a complexity analysis of the time, space, and number of messages required to elect a leader.

\end{abstract}


%
\IEEEpeerreviewmaketitle

\section{Introduction}
\label{intro}
%
%
%
%
\IEEEPARstart{T}{he} existence of a leader, i.e., a coordinator, is a necessary condition for the completion of various tasks in distributed computing \cite{zhao2002information,guo2004dynamic,eltoweissy2006dynamic}.
Leader election (LE) in dynamic distributed networks is a fundamental, yet challenging, task that concerns research to this day \cite{dynamicnets2011,fernandez2017distributed}. 
State of the art leader election algorithms for distributed dynamic networks are based on strong assumptions on either synchronization of nodes / communication or network stability. 
Previous research utilized distributed synchronization to minimize the synchronization assumptions but continued to place strong limits on the frequency of possible network changes.%
\footnote{Moreover, distributed message-based synchronization results in overhead of at least \(O(E)\) messages per round, where E is the number of communication links in the network~\cite[Chapter 11, page 247]{attiya2004distributed}.}  
Overall, the assumptions required in order to prove the correctness of leader election algorithms limit the applicability of these algorithms in very dynamic (agile) low-cost distributed systems such as sensor networks. 
Determining the weakest assumptions regarding synchronization and the stability of the network for leader election is an open problem \cite{Larrea2011}.

In this paper we propose a leader election algorithm for partially asynchronous networks~\cite{tseng1990partially} that allows an unbounded frequency of node failures and recoveries. 
The following are the key properties of the proposed Partially Asynchronous Agile Leader Election (PALE) algorithm:
\newline\noindent(1) Agile: In contrast to those leader election algorithms that require a period of stability in order to terminate, PALE allows any number of nodes to join or leave the network continuously during the algorithm’s execution. 
PALE even tolerates cases where the strongest nodes jitter, i.e., fail and recover, at the most inappropriate times.
\newline\noindent(2) Partially asynchronous: We assume reliable broadcast with a known upper bound on the message delay but an arbitrary message order; and a known upper bound on the drifts of the node clocks, regardless of the clocks' offsets.
To the best of our knowledge, PALE achieves the best trade-off between synchronization and stability assumptions among state of the art algorithms.

\subsection{PALE in a Nutshell}
\label{sec:nutshell}
The following are the major highlights of PALE, the proposed Partially Asynchronous Agile LE algorithm.
PALE is based on the main principle of the bully algorithm where strong machines are the preferred candidates for serving as the leader. 
Node ranks are initialized based on the nodes' physical properties such as CPU speed and available memory.
Unique node identities are used as tie breakers if necessary. 
Node ranks and identities are communicated using broadcast messages.

We assume reliable communication with bounded delays but arbitrary message delivery order. 
Nodes operate in asynchronous rounds defined according to their internal timers. 
These rounds may differ in duration and offsets due to discrepancies caused by the nodes' internal clocks.
Although the rates of the nodes' clocks are unknown, we assume an upper bound on the clock drifts.

Further we assume that nodes do not maintain information about their neighborhood and lack persistent memory. 
Instead, a volatile in-memory priority queue maintains IDs and ranks contained in incoming messages. 
In order to reduce the number of messages, broadcast messages are sent only by nodes that consider themselves to be the best nodes in their broadcast domain. 
An upper bound on the communication delay ensures agreement and the uniqueness of a leader when a node declares itself as such.

One of the most difficult challenges of leader election in agile environment is termination in the presence of two highly ranked jittering nodes that keep replacing each other at the leading position. 
PALE addresses this challenge by incrementing ranks of (non-jittering) nodes whenever they detect a failure of their leading node. 
Relying on the rank increments, we prove finite time termination of PALE algorithm if there is at least one node in the broadcast domain that lives long enough to survive the algorithm's execution.

The rest of the paper is structured as follows. 
In Section \ref{sec:related} we discuss and summarize work related to this study. 
Section \ref{sec:def-and-ass} presents our definitions and assumptions. 
In Section \ref{sec:thealg} we provide the pseudo-code and a detailed description of the leader election algorithm. 
Following the formal description, in Section \ref{sec:correctness} we prove agreement, uniqueness, and finite time termination. 
Section \ref{sec:complexity} presents a complexity analysis of the time, space, and number of messages. 
In Section \ref{sec:extensions} we outline three possible enhancements of the proposed algorithm: merging networks, self-stabilization, and adaptation to wireless ad hoc networks. 
Finally, Section \ref{sec:conc} summarizes and concludes the paper.

\section{Related Work}
\label{sec:related}

Numerous algorithms exist in the literature for distributed leader election in static, as well as dynamic networks, with different settings \cite{attiya2004distributed}. 
We begin with a description of the most relevant characteristics of existing leader election algorithms: 

\textbf{Tolerance to changes --} In the majority of the related work distributed leader election algorithms are classified as either static (do not tolerate changes to the network during algorithm execution) or dynamic (do tolerate changes). 
In this study we consider a finer classification: \emph{static}, \emph{dynamic}, and \emph{agile} algorithms.

\emph{Static algorithms.} 
Do not tolerate changes in the network during the algorithm's execution, i.e., one cannot prove termination, agreement, and uniqueness.

\emph{Dynamic algorithms.}
Allow changes in the network during the algorithm's execution, i.e., the ability to prove the correctness of the algorithm is not impaired by changes. 
However, termination requires a certain period of stability in the network, and thus a high enough frequency of changes prevents the algorithm from terminating. 
Consequently, there are limitations regarding the possible changes (for example, a bounded frequency of failures). 

\emph{Agile algorithms.}
Same as dynamic but do not require a period of stability and can terminate in the presence of changes.

\textbf{Synchronization --} An algorithm may be classified based on the network synchronization level it requires: \emph{synchronous}, \emph{partially asynchronous}, or \emph{fully asynchronous} \cite{tseng1990partially,dynamicnets2011,attiya2004distributed}.

\emph{Synchronous algorithms.}
Assume that the message delay is bounded and the processes either operate at synchronous rounds (lockstep) or have access to a global clock \cite[Chapter 2, page 12]{attiya2004distributed}.
	
\emph{Partially asynchronous algorithms.}    
Only assume that the message delay is bounded \cite{tseng1990partially,dynamicnets2011}. 
Some partially asynchronous algorithms bound the message delay by bounding the dynamic network diameter (also called flooding time) \cite{dynamicnets2011}.
	
\emph{Fully asynchronous algorithms.}
Assume that either the message delay is unbounded or nodes may pause for an unlimited amount of time. 
In the latter case, it is not possible to guarantee bounds on the time between message production and consumption.

Consensus, and in particular leader election, is not possible in fully asynchronous agile networks~\cite{Fischer85}. 
Determining the weakest assumptions regarding synchronization and network changes that allow termination is still considered an open problem~\cite{Larrea2011}.

In the rest of this section we explore the trade-off between tolerance to changes and synchronization. 
We focus on asynchronous or partially asynchronous leader election algorithms that assume there are some amount of changes in the network. 
The relevant literature is summarized in Table \ref{tab:related}. 
Static and synchronized algorithms, such as~\cite{dolev1997uniform,le_manet2000,le_manet2005,rcle2010,staticAsynchLE2012,robustLE2013}, are out of the scope of this discussion.

\subsection{Dynamic algorithms in partially asynchronous networks}
This category includes algorithms with both limitations on network changes and bounded message delays, and more specifically, algorithms that construct spanning trees (or directed acyclic graphs (DAGs)) to reach consensus or to ``spread the word'' about the leader's identity \cite{le_wireless2003,le_manet2004,le_mobnet2006,datta2010self}. 
When nodes fail or join the network these algorithms must reconstruct the tree before reaching a consensus. 
Since tree reconstruction is not immediate, frequent changes in the network may prevent construction of a valid tree, and thus, prevent the algorithm's termination.

All of the above algorithms use acknowledgment (ACK) or beacon messages during the tree's construction or maintenance. 
Algorithms that rely on ACKs implicitly assume a bound on message delay, as this is required to define timeout. 
Some algorithms do not rely on the construction of trees but use ACKs to ensure the completion of one phase before moving to the next \cite{Singh2011,fernandez2017distributed}.

\citet{Singh2011} aimed at minimizing the overhead of electing a new leader when the current leader has failed. 
During the election phase a cabinet of elite nodes is elected. 
The goal of this cabinet is to appoint a replacement of a failing leader, thus, avoiding a new election process. 
If the elite nodes have also failed, a new leader election has to take place. 
Maintaining continuity of leadership while minimizing message overhead is especially important for agile, low-cost distributed systems, such as sensor networks. 
In this paper we address this objective by ensuring that only nodes that regard themselves as leading participants engage in the new leader election.

\citet{asynchLE_manet2006} propose an algorithm where a leader is elected by a majority of nodes. 
This is an example of an algorithm that does not require a spanning tree (or DAG) construction and does not rely on ACKs. 
However, the authors assume bounded point-to-point message delay and link failure detection. 
They also assume a minimum lifetime of nodes that will allow the election message to pass through the majority of nodes.

\subsection{Dynamic algorithms in asynchronous networks}
Ingram et al. \citep{asynchLE_dynnet2009,Ingram2013} construct a leader-oriented DAG in each connected component. 
However, they loosen the synchronization assumptions by replacing the bounded message delays with causal clocks and link failure detection. 
Note that the link failures are propagated through the reconstruction of the forming leader-oriented DAG.
As noted earlier this reconstruction prevents the algorithm's termination in agile networks, and the algorithm terminates only after topology changes cease. 
When the algorithm terminates happens every connected component contains a unique leader.

\citet{Larrea2011} propose another algorithm for leader election in dynamic networks that does not require an upper bound on message delay. 
However, this algorithm is based on the assumption that the network's behavior eventually becomes monotonic, i.e., nodes either only join (non-decreasing behavior) or leave (non-increasing behavior) the network. 
As a consequence, this algorithm does not operate well in presence of jittering nodes in the network.

In the same paper the authors perform an in depth investigation of the trade-off between an environment's synchronization and tolerance to changes in the network. 
As an alternative to the algorithm requiring monotonic network behavior, the authors present another algorithm that requires a moment of stability in the network. 
In this case (according to the new requirement), in order for the leader to be elected and not be demoted until it crashes or leaves the network, there should be a time \(t\) at which none of the messages currently in transit has been sent by a node that has left the network before time \(t\). 
This, in turn, means that no nodes have failed for at least the amount of time of the \emph{message delay}. 
The authors state that an upper bound on the message delay and a bound on the differences between nodes' clocks are required to prove termination. 
We categorize this alternative as dynamic and partially asynchronous algorithm.

As opposed to \citet{Larrea2011}, we do not require monotonic behavior or a moment of stability but allow any number of nodes to join or leave the network at any moment during the algorithm's execution. 
Similar to the second algorithm proposed by \citet{Larrea2011}, we rely on bounded message delays (a feasible requirement in real systems). 
Next we elaborate on partially asynchronous algorithms that attempt to weaken the assumptions regarding network changes.

\subsection{Agile algorithms in partially asynchronous networks}
\citet{Melit2011} propose a partially asynchronous leader election algorithm that allows some nodes to join or leave the network during the algorithm's execution. 
The authors state that in order to terminate, their algorithm requires a finite number of topology changes. 
However, we consider this algorithm as agile, because the above assumption seems to be redundant, and their algorithm only requires that the node with the lowest identifier is not demoted eventually. 
In any case, the node with the lowest identifier cannot enter and leave the network repeatedly, i.e., cannot jitter.

The optimal regional consecutive leader election algorithm (Optimal RCLE)~\cite{optimalrcle2011} assumes a fixed geographic region with bounded communication diameter and unique IDs. 
Before engaging in the leader election a node enters a waiting phase. 
A node exits the waiting phase when it becomes the oldest node and competes for leadership or when a leader is chosen. 
Similar to the proposed PALE algorithm, Optimal RCLE assumes that there is at least one node stable enough for the algorithm to terminate and allows all other nodes to leave or join the network at arbitrary times.
However, (1) in Optimal RCLE strong nodes are not preferred over weak nodes. 
(2) Optimal RCLE assumes that node clocks are running at the same speed. 
(3) In contrast to \citet{Singh2011} and the proposed PALE algorithm, in Optimal RCLE when leader failure is detected, all nodes that have exited their waiting phase compete for leadership, resulting in higher message overhead.

Similar to \citet{optimalrcle2011} and \citet{fernandez2017distributed} we penalize unstable nodes and prefer veteran nodes that have remained in the network for a period of time. 
Termination proof relies heavily on the preference of veteran nodes in all three of these algorithms, however PALE manages the trade-off between stability and the physical parameters of nodes through its adaptive rank formula (see Equation~\ref{eq:rank} in Section \ref{sec:rank}). 

\subsection{Agile algorithms in asynchronous networks}
\citet{Fischer85} prove that given asynchronous communication, consensus is not possible in the presence of failures. This impossibility result can also be obtained by combining the results of \citet{Dolev1987} and \citet{WELCH1987159}. 
\citet{Larrea2011} discuss the trade-off between asynchronous communication and tolerance to network changes with respect to leader election algorithms. 
In the previous subsections we explore this trade-off from different perspectives through our review of the assumptions made in the literature.

\begin{table*}
	\caption{The Two Dimensions of Leader Election}
	\label{tab:related}       
		\begin{tabular}{cc|P{2.9in}|P{2.5in}|}
			\cline{3-4}
			& & \multicolumn{2}{ c| }{Synchronization} \\ 
			\cline{3-4}
			& & Partially Asynch & Asynch \\ 
			\hline
			\multicolumn{1}{ |c }{\multirow{3}{*}{\rotatebox[origin=c]{90}{\parbox{1.15cm}{Network Changes}}}} &
			\multicolumn{1}{ |c| }{Dynamic} & 
			Vasudevan et al. \citep{le_wireless2003,le_manet2004}, \citet{le_mobnet2006}, \citet{asynchLE_manet2006}, \citet{datta2010self}, \citet{Singh2011}, \citet{fernandez2017distributed}& Ingram et al. \citep{asynchLE_dynnet2009,Ingram2013}, \citet{Larrea2011}\\ \cline{2-4}
			\multicolumn{1}{ |c  }{}                        &
			\multicolumn{1}{ |c| }{Agile} &  \citet{Melit2011}, \citet{optimalrcle2011} & Impossibility result by \citet{Fischer85} \\ \cline{1-4}
		\end{tabular}
\end{table*}

\section{Definitions and Assumptions}
\label{sec:def-and-ass}
In this section we provide the assumptions behind the newly proposed leader election algorithm and the definitions used in the rest of the paper. 
The following terms and assumptions are provided either from the perspective of a single node (denoted by local) or from the perspective of the network as a whole (denoted by global).

\begin{definition} [Active node (local)]
	A node is said to be \emph{active} if it can communicate (send and receive messages).
\end{definition}

\begin{definition}[Region (global)]
	A region \(R\) is a set of active nodes within the same broadcast domain (communication range).
\end{definition}
In this research we focus on LE in a single region. 
The maximal possible number of nodes in a region is denoted by \(n=|R|\).

A node engages in the LE process each time it becomes active within a region or when it is disconnected from its current leader. 
When a node gets disconnected from the network or fails (becomes \emph{inactive}), it stops executing the LE process. 
Then, when it regains connection, the node starts a new LE process without having persistent information from previous executions. 
Since nodes can become inactive repeatedly during the LE process, there is a risk that no node will survive long enough to complete the process.

\begin{assumption}[A stable node (global)]
	\label{ass:stable-node}
	At least one of the nodes in the region is stable enough in order to survive entire LE process.
\end{assumption}
Note that we do not assume knowing the identity of the stable node or its physical characteristics. 
This assumption is minimal in the sense that, if no node remains long enough in the network to be elected as a leader, then leader election is no longer relevant~\cite{Larrea2011}.

In this paper we assume that all messages reach their destinations. 
This requirement can be met using one of several well-known broadcasting algorithms \cite{kaashoek1989efficient,pagani1997reliable,melliar1990broadcast,clementi2004round,bar1992time}.
Reliable broadcasting prolongs the message delays in the network due to retransmission of messages.
According to \citet[Chapter 2, page 14]{attiya2004distributed} the \emph{delay} of a message is the time that passes between the event of initiating the message transmission by the source node and the event of consuming the message by the target node.

\begin{assumption}[Message reachability (global)]
	\label{ass:broadcast-reach}
	Every broadcast message sent by any node in a region $R$ is heard by all other nodes in the region during \(\delta_d\) which denotes the message's maximal \emph{delivery time}.
\end{assumption}

We assume that nodes operate in rounds timed according to their internal clocks. 
Let $\delta_{r_v}$ denote the length of \(v\)'s round. 
The length of rounds may differ slightly for different nodes due to clock drifts \cite{clockdrift2006} or scheduler discrepancies. 
We assume that the clocks of the nodes in the network may run at different rates. 
But the maximal ratio between clocks' rates, and hence, the maximal ratio between nodes' rounds, is bounded.

\begin{assumption}[Maximal rounds ratio (global)]
	\label{ass:clocks}
	The fastest round in the region is shorter than the slowest one by at most a factor of \emph{MaxRatio}:
	\[\forall_{u,v}, \frac{\delta_{r_u}}{\delta_{r_v}}\leq MaxRatio\]
\end{assumption}
Note that this assumption is weaker than the common practice in LE algorithms. 
Most solutions assume that the clock rates are the same (for example, \cite{optimalrcle2011}) or rely on distributed synchronization mechanisms (for example, \cite{Singh2011,fernandez2017distributed}).

At the beginning of each round, each node may broadcast a message, update its rank, or declare itself as a leader (see \emph{On RoundTimer timeout} in Algorithm~\ref{alg:le}). 
Let \(\delta_{OnTimer_v}\) denote the time it takes \(v\) to perform these operations. 
During the rest of the round a node handles received messages. 
Let \(\delta_{OnMsg_v}\) denote the time it takes \(v\) to handle a single received message.

In order to ensure that the proposed algorithm works properly (and that incoming messages do not pile up), we need to ensure that the length of the shortest round is sufficiently long.

\begin{assumption}[Round length (local)]
	\label{def:shortest-round}
	We assume that the \emph{round} length of all nodes is greater than the amount of time required for all messages to be sent, delivered, and processed.	
	\[
	\forall_{u,v}
	\delta_{r_v} \geq \delta_{OnTimer_u} + \delta_d + \delta_{OnMsg_v}\cdot n \cdot MaxRatio 
	\]
\end{assumption}


\begin{definition}[Leading Participant (local)]
	A node \(v\) considers a node \(u\) as the \emph{leading participant} if \(u\) has reported the highest rank among all nodes that \(v\) has received a message from up until its latest round, and \(u\)'s rank is higher than \(v\)'s own rank.
	
\end{definition}
By \emph{rank} we refer to a quantity that encapsulates the physical characteristics of a node, e.g., CPU, physical memory, network bandwidth, etc., and its stability. 
We elaborate on the computation of rank in Section~\ref{sec:rank}.   

\begin{definition}[Handshake (local)]
	We say that a node \(v\) has elected a leading participant \(u\) as a leader and become its ``slave'' if and only if \(v\) performed a handshake with \(u\), denoted by \(handshake(v,u)\).
\end{definition}
In our implementation, a handshake (Algorithm~\ref{alg:handshake}) is performed by establishing a direct and reliable communication channel, such as a TCP connection between a node and its leader.

\section{The Proposed Algorithm}
\label{sec:thealg}
In this section we describe an LE algorithm which can be applied to dynamic distributed networks. 
The network is composed of nodes with various capabilities that can join and leave the network at will, without any persistent knowledge about the nodes in the region. 
Two nodes can communicate and exchange messages if they are in the same region. 
The clocks of the nodes are not synchronized. 
Moreover, the clocks may run at different speeds while in compliance with Assumption \ref{ass:clocks}.

The main design considerations in the development of the algorithm are: 
(1) minimizing the number and size of the messages; 
(2) minimizing the memory and CPU usage of the competing nodes; and 
(3) electing a suitable (stable and strong enough) node as a leader within a finite amount of time.

As noted in the previous section, rank is a quantity that encapsulates the physical characteristics of a node and its stability. 
Therefore, a flexible rank formula was designed, so that the rank of a stable node grows as long as it is active. 
This property will prevent a situation in which a strong but jittering node prevents a stable but less strong node from being elected as a leader.

\subsection{Computing the Rank of a Node}
\label{sec:rank}
The rank of a node is based on its physical properties (quantified by physScore) and the node's ability to continuously remain active (quantified by stabilityCounter). 
The StabilityCounter value increases while the node is active and resets when the node becomes inactive.
To ensure the correctness of the algorithm, a stable node must eventually accumulate a rank higher than the maximally possible physScore value.

\begin{equation}
\label{eq:rank}
rank = w\cdot stabilityCounter + physScore
\end{equation}
Equation \ref{eq:rank} presents the formula used to calculate the rank, which is used in the proposed algorithm.

In order to meet the requirement stated above, we suggest increasing the rank linearly with stabilityCounter. 
The growth rate, \(w\), of the rank should be low enough to allow nodes with a high physical score (physScore) to be elected. 
Note that if two nodes have the same rank, their IP or MAC addresses can be used to break the tie.

The flexible design of the algorithm provides the ability to manipulate the LE process by modifying the formula used to calculate the rank in order to elect a leader that best corresponds to the objectives.
Note that the formula must ensure that the score of active nodes increases over time.

\subsection{Formal Description of the Algorithm} 
\label{sec:formal-desc}

\algrenewcommand\algorithmicrequire{\textbf{Input:}}
\algrenewcommand\algorithmicprocedure{\textbf{event}}

\begin{algorithm*}
\caption{Leader Election}
\label{alg:le}
\begin{algorithmic}[1]
	\Require $maxRound$; $msgDeliveryTime$; $maxRatio$; $w$
	\Procedure{Initialization}{}	
		\State $PL \gets \{\text{current node}\}$ \Comment{a sorted list of the LE participants}
		\State $cntRounds \gets 0$ \Comment{num of rounds that the current node is active}
		\State $roundsAsLeading \gets 0$ \Comment{num of successive rounds that the current node is the leading participant}
		\State $pl0DelCnt \gets 0$ \Comment{num of times that the current node lost connection with a leading participant}
		\State $lastLeadMsg \gets 0$ \Comment{the last round in which the current node received a BeepMsg from the leading participant}
		\State $IamLeader \gets False$
		\State $rank \gets \text{ComputeRank}\left(w, pl0DelCnt\right) $
		\State $\text{Broadcast}\left(\text{BeepMsg}\left(\text{now()}, rank, IP, roundsAsLeading\right)\right)$
		\State $\text{StartTimer}(RoundTimer)$
	\EndProcedure
	\Statex
	\Procedure{RoundTimer timeout}{}\label{alg:le:roundtimer}	
		\If{$IamLeader \equiv True$}
			\State $\text{Broadcast}\left(\text{BeepMsg}\left(\text{now()}, rank, IP, roundsAsLeading\right)\right)$
		\Else
			\State $cntRounds++$
			\If{$PL[0]\neq \text{current node} \wedge cntRounds-lastLeadMsg>maxRatio $}
				\State $PL[0].delete()$
				\State $pl0DelCnt++$
				\State $rank \gets \text{ComputeRank}\left(w, pl0DelCnt\right)$
				\State $PL.\text{insertOrUpdate}\left(\text{BeepMsg}\left(\text{now()}, rank, IP, 0\right)\right)$
			\EndIf
			\If{$PL[0]\equiv \text{current node}$}
				\State $roundsAsLeading++$		
				\If{$roundsAsLeading\equiv maxRound$}
					\State \(rank\gets\infty\) \label{alg:le:infty}
					\State \(IamLeader\gets True\)
				\EndIf
				\State $\text{Broadcast}\left(\text{BeepMsg}\left(\text{now()}, rank, IP, roundsAsLeading\right)\right)$
			\EndIf
		\EndIf	
	\EndProcedure
	\Statex
	\Procedure{vBeepMsg received}{$v$}\label{alg:le:beepmsg}
	\If{$PL[0]\equiv v \wedge PL[0].round > vBeepMsg.round \wedge PL[0].time < vBeepMsg.time$ }
		\State $PL[0].delete()$
		\State $pl0DelCnt++$
		\State $rank \gets \text{ComputeRank}\left(w, pl0DelCnt\right)$
		\State $PL.\text{insertOrUpdate}\left(\text{BeepMsg}\left(\text{now()}, rank, IP, 0\right)\right)$
	\EndIf
	\If{$PL[0]\equiv \text{current node} \wedge v\neq \text{current node} \wedge rank < vBeepMsg.rank$}
		\State $roundsAsLeading \gets 0$
	\EndIf
	\State $PL.\text{insertOrUpdate}\left(vBeepMsg\right)$
	\If{$PL[0]\equiv v \wedge vBeepMsg.round \geq maxRound$}
		\State $\text{Handshake}\left(\text{current node}, v\right)$ \label{alg:le:handshake}
	\EndIf
	\If{$PL[0]\equiv v$}
		\State $lastLeadMsg\gets cntRounds$
	\EndIf
	\EndProcedure
\end{algorithmic}
\end{algorithm*}

\begin{algorithm*}
\caption{Broadcast}
\begin{algorithmic}[1]
	\Require $msg$; $numOfCopies$ -- the number message's copies the node sends.
	\State $cntCopies \gets 0$
	\For{$cntCopies < numOfCopies$}
		\State	Broadcast($msg$)
		\State $cntCopies++$
	\EndFor
\end{algorithmic}
\end{algorithm*}

\begin{algorithm*}
	\caption{ComputeRank}
	\label{alg:rank}
	\begin{algorithmic}[1]
		\Require $w$; $pl0DelCnt$
		\Statex Retrieve Windows Experience Index assessments of key system components.
		\Statex Each score ranges between 1.0 and 7.9.
		\State $ramScore \gets \text{getRAMScore()}$
		\State $processorScore \gets \text{getProcessorScore()}$
		\State $hardDiskScore \gets \text{getHardDiskScore()}$
		\Statex The aggregated score also ranges between 1.0 and 7.9.
		\State $physScore \gets \text{aggregate}\left(ramScore, processorScore, hardDiskScore\right)$
		\State $physScore \gets \frac{physScore}{7.9}$
		\State $rank \gets w\cdot pl0DelCnt + physScore$
		\State \textbf{return} $rank$
	\end{algorithmic}
\end{algorithm*}

\begin{algorithm*}
	\caption{Handshake}
	\label{alg:handshake}
	\begin{algorithmic}[1]
		\Require $v$
		\If{Not Connected to $v$}
			\State Establish connection with $v$
		\Else
			\State Ignore		
		\EndIf
	\end{algorithmic}
\end{algorithm*}

Next we describe the pseudo-code of the proposed LE algorithm (Algorithm~\ref{alg:le}) from the perspective of a single node \(u\). 
Each node \(u\) maintains a sorted list of participants from which it received a message. 
When a node \(u\) joins the region it performs the following steps: 
initializes its $PL$ list (a sorted list of the LE participants) such that $PL[0]$ points to \(u\) (Line 2); 
initializes the $cntRounds$ (number of rounds that the node is active), $roundsAsLeading$ (number of rounds that the node is a leading participant), and $pl0DelCnt$ (number of times that the current node lost connection with the leading node) counters (Lines 3, 4, and 5, respectively); 
initializes $lastLeadMsg$ (the last round in which the current node received a broadcast message from the leading node) parameter (Line 6); 
sets $IamLeader$ to $false$ (Line 7); 
computes and broadcasts its initial rank (Lines 8-9); 
and starts the round timer $RoundTimer$ (Line 10). 
Upon the timeout of the round timer, if the node did not receive a broadcast message from other nodes, or only received messages from nodes with ranks lower than its own, then the $PL[0]$ still points to \(u\).
In such a case, the node will increase the $roundsAsLeading$ (Line 24) and notify all nodes in the region.
The notification message includes the node's rank and its current value of $roundsAsLeading$ (Line 29).

\begin{definition}[Leader (local)]
	\label{def:leader-local}
	A node considers itself to be a leader if it considers itself a leading participant ($roundsAsLeading$) for consecutive \(MaxRounds=2\cdot\lceil MaxRatio\rceil+2\) rounds.
\end{definition}

In cases in which a node \(u\) considers itself the leading participant for $maxRounds$ rounds, the rank is set to infinity (higher than any other rank), and $IamLeader$ is set to $true$ (Lines 23-30). 
The leader will continue broadcasting its rank every round (Line 13-14). 
From this point on, all nodes in the region will perform a handshake with the leader (\(u\)) (Lines 44-46).

During the leader election process, if node \(u\) received a broadcast message from another node \(v\) with a rank higher than its own, then $PL[0]\neq u$ holds (Lines 34-46, explanation follows) and \(u\) updates the parameter $lastLeadMsg$ with the value of $cntRounds$ (Lines 47-49).

Upon a message $m$ from some other node $v$, the current node ($u$) checks whether it loses the leadership to $v$ (Line 40). 
If so, the current node resets the $roundsAsLeading$ counter (Line 41). 
The current node adds $v$ to $PL$ in a sorted manner according to \(v\)'s rank in the message (Line 43).

If the node $v$ has been the leading participant for at least $maxRound$ rounds (Line 44), the current node initiates a handshake -- direct communication channel -- with $v$ (Lines 44-46). 
If the leading node $PL[0]$ is the one that $u$ has received message $m$ from (i.e., $v$), $u$ sets $lastLeadMsg$ to the current round index (Lines 47-49). 
New nodes that join the region engage in the leader election process for at most \(\lceil MaxRatio \rceil + MaxRounds\) rounds until they perform a handshake with the leader: \(\lceil MaxRatio \rceil\) rounds until they receive a message from the leader and another \(MaxRounds\) rounds in order to ensure that he is indeed the leader.

A node that is not the leading participant should occasionally receive a message from the leading participant. 
After waiting for the $\lceil MaxRatio\rceil$ rounds, the node checks whether it should have received a message from the current leading participant (Line 17). 
If more than $\lceil MaxRatio\rceil$ rounds have passed and a message from the $PL[0]$ has not been received, it means that the $PL[0]$ is currently unavailable (otherwise, a message from that node would already have been received). 
Therefore, the node removes $PL[0]$, increases the $pl0DelCnt$ and recomputes its rank, and updates its location in the $PL$ list (Lines 18-21).

Lines 34-39 address the extreme case in which the leading participant of the node $u$ fails and recovers between two consecutive $RoundTimer$ timeout events of $u$. 
Upon the receipt of a $BeepMsg$ from $v$, if the previous $PL[0]$ has also been $v$, $u$ verifies that the current $BeepMsg.round$ is greater than the $round$ of the previous message from $v$. 
The message order is determined by comparing the local timestamps of $v$.

In cases in which the round number in the newer message is lower than the round number stored in \(PL[0]\), we conclude that $v$ has failed. 
In such a case, $u$ performs the same set of actions as in Lines 18-21: 
(1) deletes $PL[0]$ -- $v$'s already irrelevant record, 
(2) increases the counter $pl0DelCnt$, 
(3) updates its own rank, and 
(4) updates the $PL$ list. 

To ensure the correctness of the implementation, \emph{On RoundTimer timeout} (Line \ref{alg:le:roundtimer}) and \emph{On vBeepMsg received} (Line \ref{alg:le:beepmsg}) blocks of code should be atomic.

\begin{figure}
	\centering
	\includegraphics[width=0.6\textwidth, clip, trim=270 50 150 50]{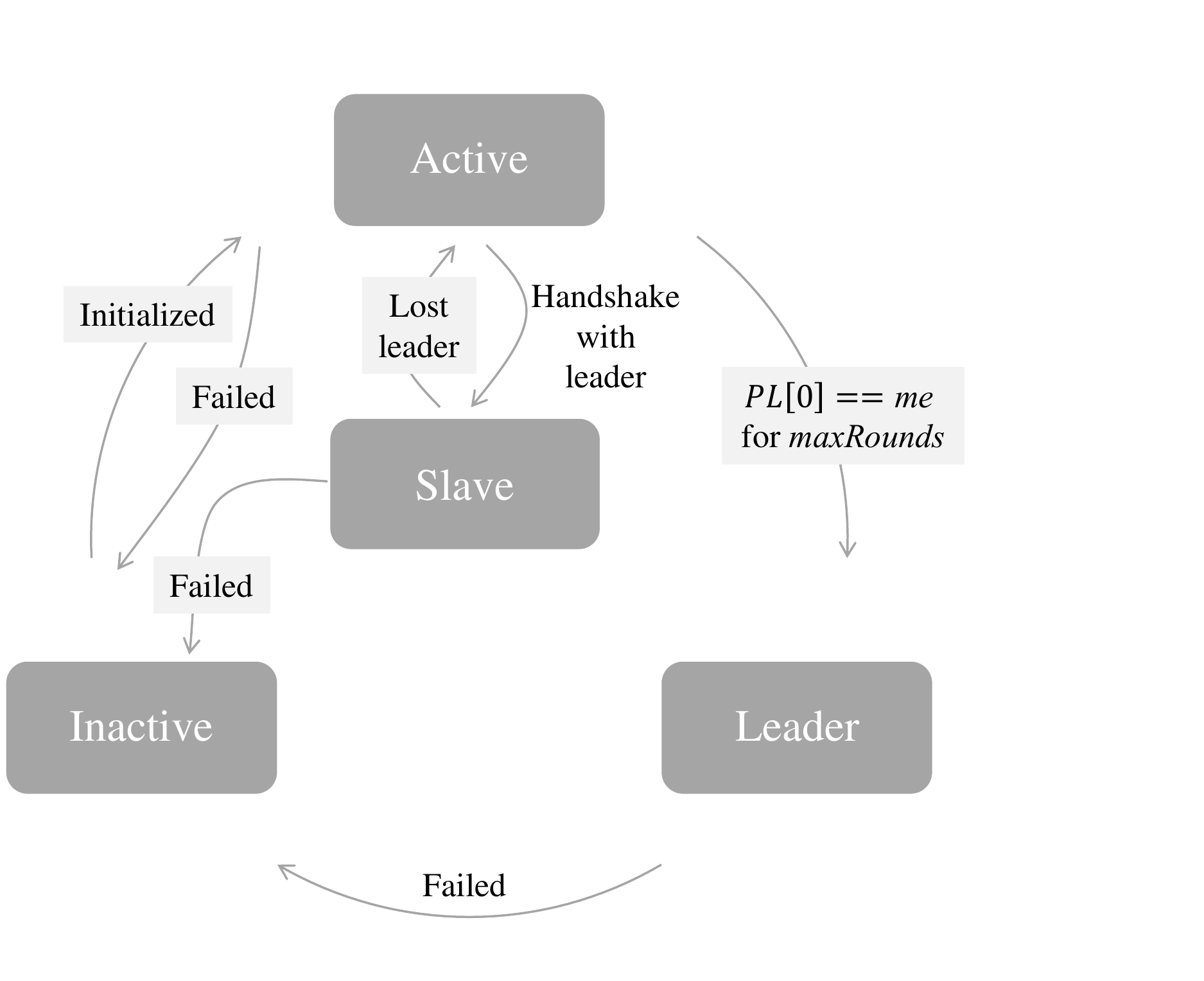}
	\caption{\label{fig:lifecycle}
		Node's Life Cycle}
\end{figure}

Figure \ref{fig:lifecycle} illustrates the life cycle of a single node in a region where the LE algorithm is executed.

\section{Proof of the Algorithm's Correctness}
\label{sec:correctness}
To prove the correctness of the LE algorithm we must prove that the following conditions hold:
\begin{enumerate}
	\item Agreement -- All nodes in the region that elect a leader elect the same leader.
	\item Uniqueness -- There is exactly one node that considers itself a leader.
	\item Termination -- If some active node \(v\) remains in the region for a sufficiently long period of time, \(v\) must reach a state where it has a valid leader.
\end{enumerate}

\begin{definition}[Termination (global)]
	\label{def:term-global}
	We say that a \emph{leader}'s identity is determined if and only if at least one node in a region has become a leader and is ready to accept handshakes (\(IamLeader == true\)).	
\end{definition}

\begin{definition}[Agreement (global)] 
	\label{def:agree}
	At any point in time, the \emph{agreement} condition is satisfied if and only if all nodes in the region \(R\) that performed a handshake performed it with the same node \(v\in R\).
	\[
	\forall_{u,v\in R}\exists_{x,y\in R} handshake(u,x) \wedge handshake(v,y) \rightarrow x=y
	\]
\end{definition}

\begin{lemma}[Message handling delay]
	\label{lem:handle}
	Let \(u,v\) be two active nodes, \(m\) be a message that was broadcast by \(u\) at \(t_s\) and handled by \(v\) at \(t_r\) (all times according to \(u\)'s internal clock), then \(u\) can run at most \(MaxRatio+1\) rounds between \(t_s\) and \(t_r\).
\end{lemma}
\begin{IEEEproof}
	According to Assumption \ref{def:shortest-round}, the message \(m\) will reach node \(v\) before \(t_s\) plus one additional round. 
	$v$ will handle $m$ during the round in which $m$ was received.
	
	In addition, according to Assumption~\ref{ass:clocks}, $u$ can run at most $MaxRatio$ additional rounds before $v$ finishes handling $m$. 
	Thus, $u$ can run at most $MaxRatio+1$ rounds between $t_s$ and $t_r$.
\end{IEEEproof}

\begin{lemma}[Upper time limit between receiving consecutive messages]
	\label{lem:minMaxround}
	Given a leading participant $v$ and two consecutive beep messages $m_1,m_2$ emanating from $v$, any node $u$ may run at most \(2\cdot MaxRatio + 1\) rounds between arrivals of $m_1$ and $m_2$.
\end{lemma}
\begin{IEEEproof}
	Let us denote the rounds at which $v$ sends $m_1$ and $m_2$ as $i$ and $i + 1$, respectively. 
	In the worst case, $v$ sends $m_1$ at the beginning of round $i$ and $m_2$ at the end of round $i + 1$. 
	Therefore, according to Assumption~\ref{ass:clocks}, $u$ will run at most \(2\cdot MaxRatio\) rounds between the transmissions of $m_1$ and $m_2$. 
	These rounds include the round during which $u$ handles $m_1$. 
	Finally, at most another round of $u$ will be needed in order to receive and handle $m_2$.
\end{IEEEproof}

\begin{theorem}[Uniqueness of a leader]
	\label{lem:uniqueness}
	At any point in time only a single node in a region $R$ may accept handshakes (i.e., consider itself a leader).
\end{theorem}
\begin{IEEEproof}
	According to Lines 25-28, a node considers itself a leader only if its \(roundsAsLeading\) counter has reached \(maxRound\) in accordance with Definition~\ref{def:leader-local}. 
	We will prove that two nodes $u$ and $v$ (\(u\neq v\)) cannot run consecutive \(maxRound=2\cdot MaxRatio+2\) rounds, considering themselves leading participants simultaneously.
	
	Assume in contradiction that there are two nodes \(u\) and \(v\) that run consecutive \(maxRound=2\cdot maxRatio+2\) rounds (according to their internal clocks), considering themselves leading participants (\(v.PL[0]=v \wedge u.PL[0]=u\)). 
	For the rest of this proof we will only address time from the perspective of \(u\)'s internal clock.
	
	Let \(t^u_s\) and \(t^v_s\) denote the round start time (Line~\ref{alg:le:roundtimer} in Algorithm~\ref{alg:le}) of the nodes \(u\) and \(v\), respectively, when they become leading participants. 
	Let \(t^u_e\) and \(t^v_e\) denote the round start time of the nodes \(u\) and \(v\), respectively, when they declare themselves leaders.	 
	Please refer to Figure~\ref{fig:lifecycle} for the node life cycle. 

	\begin{figure*}
		\centering
		\includegraphics[width=\textwidth, clip, trim=0 5 0 0]{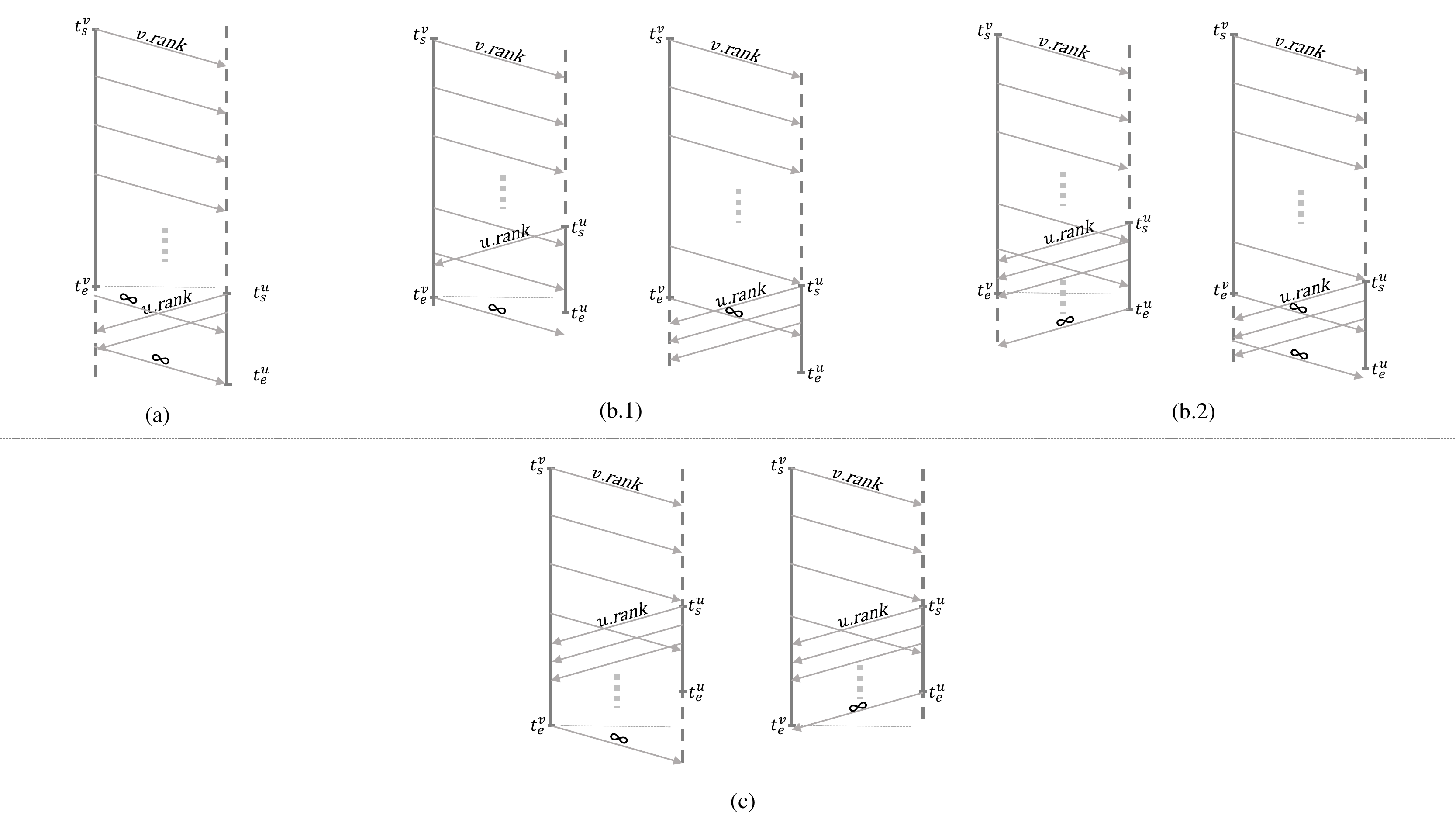}
		\caption{\label{fig:uniqueness}
			Message sequences in the cases: (a) $t^v_e\leq t^u_s$; (b.1) $t^u_s < t^v_e < t^u_e$ and $v.rank > u.rank$; (b.2) $t^u_s < t^v_e < t^u_e$ and $v.rank < u.rank$; (c) $t^v_e \geq t^u_e$}
	\end{figure*}

	Assume without loss of generality that \(t^v_s \leq t^u_s\).
	Consider the following cases (see Figure~\ref{fig:uniqueness}): 
	\begin{enumerate}
		\item \(t^v_e\leq t^u_s\) (see Figure \ref{fig:uniqueness}(a)):
		
		$v$ became a leader (according to Definition~\ref{def:leader-local}) and set its rank to infinity before or at the same time that $u$ became a leading participant. 
		$v$ continues broadcasting messages $m_L$ with the updated rank (\(v.rank=\infty\)).
		
		In the worst case, each round of $v$ is equal to $maxRatio$ rounds of $u$ (see Assumption~\ref{ass:clocks}). 
		Therefore, $v$ will broadcast $m_L$ at most $maxRatio$ rounds after \(t^u_s\) according to $u$'s internal clock. 
		According to Assumption~\ref{def:shortest-round}, this message will reach $u$ before the end of round $maxRatio+1$ after \(t^u_s\).
		Therefore, $u$ will receive the message from $v$ with the updated rank at least once during \(maxRound=2\cdot maxRatio+2\) rounds and cease being the leading participant in Line 43 in Algorithm~\ref{alg:le}) before \(t^u_e\).
		
		\item \(t^u_s < t^v_e < t^u_e\)
		\begin{enumerate}
			\item \(v.rank > u.rank\) (see Figure \ref{fig:uniqueness}(b.1)): \\
			Given \(maxRounds=2\cdot maxRatio+2\), node $u$ will receive a message from $v$ before \(t^u_e\) (Lemma~\ref{lem:minMaxround}). 
			This message will either contain $v$'s rank (computed in Lines 20 or 37) or infinity (Line 26). 
			In both cases, $u$ will receive a message with a rank value that is larger than $u$'s own rank. 
			Thus, $u$ will cease being the leading participant before \(t^u_e\).
			
			\item \(v.rank < u.rank\) (see Figure \ref{fig:uniqueness}(b.2)):\\
			Consider the first beep message $m_0$ broadcast by $u$ as a leading participant. 
			There are two cases: 
			\begin{enumerate}
				\item 
				$v$ receives $m_0$ before \(t^v_e\): \\
				$v$ ceases being the leading participant before \(t^v_e\).
				
				\item 
				$v$ does not receive $m_0$ before \(t^v_e\):\\
				In this case \(u\) must have sent the message later than \(maxRatio+1\) rounds before \(t^v_e\) according to Lemma~\ref{lem:handle}. 
				This means that $u$ must run at least $maxRatio+1$ between \(t^v_e\) and \(t^u_e\). 
				Let $t_s$ be the time when $v$ broadcast its first message $m_L$ as a leader (\(v.rank=\infty\)). 
				$v$ will run at most one round between \(t^v_e\) and \(t_s\), thus, $u$ can run at most $maxRatio$ rounds between \(t^v_e\) and \(t_s\). 
				$u$ can run at most one additional round after $t_s$, during which it receives and handles the message. 
				Thus, $u$ must handle $m_L$ before \(t^u_e\).
			\end{enumerate}
		\end{enumerate}
		
		\item \(t^v_e \geq t^u_e\) (see Figure \ref{fig:uniqueness}(c)):
		\begin{enumerate}
			\item \(v.rank > u.rank\):\\
			Given \(maxRounds=2\cdot maxRatio+2\), node $u$ will receive a message from $v$ before \(t^u_e\) (Lemma~\ref{lem:minMaxround}). 
			Since \(v.rank > u.rank\), $u$ ceases being a leading participant.
			
			\item \(u.rank > v.rank\):\\
			$u$ must run for at least \(2\cdot maxRatio +2\) rounds as a leading participant. 
			To the latest by the end of $u$'s first round as a leading participant $u$ will broadcast a message $m$ with its rank. 
			$u$ is faster than $v$ at most by the factor of $maxRatio$. 
			Therefore, $v$ will run at least two rounds after the message $m$ is broadcast by $u$. 
			As a result, $v$ will receive $m$ and cease being a leading participant before \(t^v_e\).
			
		\end{enumerate}
	\end{enumerate}
	We have shown that in all cases either $u$ ceases being the leading participant before \(t^u_e\) or \(v\) ceases being the leading participant before \(t^v_e\). 
	Thus, $u$ and $v$ cannot consider themselves leaders at the same time in contradiction to the initial assumption.
\end{IEEEproof}

\begin{theorem}[Agreement after termination]
	After the termination condition has been satisfied (Definition~\ref{def:term-global}), while the elected leader is active, the agreement condition (Definition~\ref{def:agree}) must hold.
\end{theorem}
\begin{IEEEproof}
	There is only one node $v$ in the region $R$ that accepts handshakes according to Theorem~\ref{lem:uniqueness}. 
	Therefore, every node that performs a handshake establishes a connection with $v$ satisfying the agreement condition (Definition~\ref{def:agree}).
\end{IEEEproof}

Finally, we prove that the termination condition holds. 
Let us begin with the intuition behind the proof. 
In the worst case, the system will contain two or more strong and jittering nodes which compete for leadership, however none of them completes $maxRound$ rounds. 
Therefore, it must be shown that the stable node $v$ (see Assumption~\ref{ass:stable-node}) will accumulate a rank higher than the rank of any other node in a finite number of steps.

According to the formula of the rank (see Equation~\ref{eq:rank}), the rank of $v$ will increase every time its leading participant fails (Lines 17-22 and 34-39). 
Eventually, in a finite number of \(v.rank \leftarrow ComputeRank(w, pl0DelCnt)\) command executions, $v.rank$ will be higher than the rank of any strong node $u$ that fails before completing $maxRound$ rounds, since $u$'s rank is initialized each time $u$ returns.

\begin{theorem}[Finite time leader election]
	\label{th:finite-time}
	The identity of a leader is determined in a finite amount of time.
\end{theorem}
\begin{IEEEproof}
	According to Assumption \ref{ass:stable-node}, there exists at least one stable node, $s$, that does not fail until a leader is elected. 
	We will prove that given $n$ nodes which periodically become active or inactive in the region, a leader will be elected in \(n\cdot k\cdot 2(maxRatio+1)^2\) rounds of node $s$, where $k$ is a constant. 
	Let us prove the required upper bound of the rounds by induction.
	
	First let us prove for \(n=2\), where one of the nodes is the stable node $s$, and the second node is denoted as $v$. 
	In the worst case, $v$'s rank is higher than $s$'s, and it repeatedly fails just before completing \(2\cdot maxRatio +2\) rounds as a leading participant. 
	If $v$ stays inactive for \(2\cdot maxRatio +2\) rounds of $s$, $s$ would become the leader. 
	Each time that $v$ fails and revives $s$ increases $s.pl0DelCnt$. 
	Thus, according to Equation \ref{eq:rank}, after $v$ fails \(k=\frac{v.physScore - s.physScore}{w}\) times, $s$'s rank will become higher than $v$'s, allowing $s$ to be elected as leader. 
	Overall this process can take up to 
	\(k\cdot (maxRatio\cdot(2\cdot(maxRatio+2)+2\cdot maxRatio+2)=k\cdot 2\cdot(maxRatio+1)^2\)
	rounds of $s$.
	
	Next, assume that there are $n−1$ nodes in the region that elect a leader at most in \((n-2)\cdot k'\cdot 2\cdot(maxRatio+1)^2\) rounds. 
	If $s$ becomes the elected leader, it means that $s$'s rank has grown above the ranks of the rest of the $n−2$ nodes (hereafter denoted as $U$).
	
	Now let us prove that given a set of $n$ nodes denoted as $V$, where \(V = \{s\}\cup U\cup \{v\}\) and \(v\neq s\), a leader will be elected in at most \((n-1)\cdot k''\cdot 2 \cdot (maxRatio+1)^2\) rounds
	(according to $s$'s internal clock), where \(k'' = k' + \frac{v.physScore - s.physScore}{w}\).

	Let us assume that there exists some node $v$ (that is the $n$th node in the region), and $v$'s rank is higher than $s$'s. 
	If $v$'s rank is higher than $s$'s, and $v$ does not fail, then in at most \(maxRatio \cdot (2\cdot maxRatio+2)\) rounds of node $s$, $v$ will be elected as leader. 
	In cases in which $v$ never completes $maxRound$ consecutive rounds, after it fails \(\frac{v.physScore - s.physScore}{w}\) times $s$'s score will be higher than $v$'s. 
	Note that the rank of any node \(u\in U\) remains lower than $s$'s rank due to the linearity of the rank. 
	Consequently, after at most another \(\frac{v.physScore - s.physScore}{w}\cdot 2(maxRatio+1)^2\) 	rounds of $s$, $s$ will be elected as a leader. 
	In total, the upper bound on the required number of rounds is \((n-2)\cdot k'\cdot 2\cdot(maxRatio+1)^2 +\frac{v.physScore - s.physScore}{w}\cdot 2(maxRatio+1)^2\). 
	Since \(k''> k'\), it can be concluded that given $n$ nodes in the region, after at most \(k''\cdot (n-1) \cdot 2\cdot (maxRatio+1)^2\) rounds the stable node $s$ will be elected as leader.
\end{IEEEproof}

\section{Complexity}
\label{sec:complexity}

In this section we analyze message, space, and time complexities.

\noindent\textbf{Message complexity.}

\noindent\emph{Harsh conditions (the worst case) --} 
In this case we consider that a higher force causes the nodes to fail at the most inappropriate times. 
In Theorem~\ref{th:finite-time} we prove that given \(n\) nodes in region \(R\), the upper bound on the number of rounds until a leader is elected is \((n-1)\cdot c\cdot 2(MaxRatio+1)^2\), where \(c\) is a constant.

Consider some fraction \(p_{join}\leq 1\) of nodes joining (or rejoining) the region in every round. 
In the worst case, each of these \(p_{join}\cdot n\) nodes broadcasts \(maxRatio\) \(BeepMsg\)s before receiving a message from a stronger participant. 
The total number of messages sent per round is $O(n)$, including messages sent by the rest, stable, \((1-p_{join})\cdot n\) nodes. 
Therefore, \(n\cdot (n-1)\cdot c\cdot 2(maxRatio+1)^2\) is an upper bound on the number of messages broadcast until a leader is elected. 
Consequently, in the worst case, the message complexity of the proposed LE algorithm is \(O(n^2)\).

\noindent\emph{Monotonic behavior --} 
Following the terminology used by \citet{Larrea2011}, we consider two types of monotonic behavior: non-decreasing (nodes only join) and non-increasing (nodes only leave). 
The extreme non-decreasing behavior of nodes is such that the nodes join gradually, ordered by their rank, where the strongest node joins last. 
When the joining node has a higher rank than the current leading participant, it becomes the new leading participant and may send additional \(maxRound-1\) messages before it becomes a leader (or a stronger node joins).

Similarly, in the extreme non-increasing behavior of nodes, the nodes leave gradually, ordered by their rank, where the strongest node leaves first. 
When the leading participant leaves the region, the next leading participant may send additional \(maxRound-1\) messages before it becomes a leader (or fails). 

In both cases, each of the $n$ nodes considers itself a leading participant and sends at most $maxRound−1$ messages. 
Therefore, the overall number of messages is $O(n)$. 
Thus, the amortized number of messages per round is $O(1)$. 

\noindent\emph{Mild conditions --} 
In this case we consider an eventually stable network. 
Whenever a leader fails, only the node that considers itself the new leading participant sends \(maxRound\) messages resulting in $O(1)$ message complexity.

Table \ref{tab:msg-complex} provides a summary of the message complexity of the three cases.

\begin{table*}
	\centering
\caption{Message Complexity Summary}
\label{tab:msg-complex}
	\begin{tabular}{| P{1.8in} |P{1.1in}|P{1.1in}|P{1.1in}|}
		\hline 
		Worst case & Monotonic behavior & Leader failure \\
		\hline 
		\(O(n)^2\) & \(O(n)\) & \(O(1)\) \\
		\hline
	\end{tabular}
\end{table*}

\noindent\textbf{Space complexity.}
Every node $v$ that participates in the LE process stores at most one object in its $PL$ list for each of the $n$ nodes in the region. 
Each object associated with some node $u$ contains $u$'s ID, $u$'s rank, the local time, and $roundsAsLeading$ (number of rounds that $u$ is a leading participant). 
Each node ID takes at most $O(\log n)$ bits, and each node stores at most $n$ objects; thus, the space complexity is $O(n\log n)$ bits.
	
\noindent\textbf{Time complexity.}
Although, most of articles describing leader election algorithms do not report the time complexity per round. 
We feel that an evaluation of the worst case time complexity is important for setting the $RoundTimer$ timeout.
	
Algorithm~\ref{alg:le}'s bottleneck is the $PL$ data structure. 
In order to reduce the complexity of the algorithm, we propose using a hybrid data structure based on \emph{Fibonacci heap} \cite{cormen2009introduction} and a \emph{hash table}. 
This will provide us the ability to: 
(1) retrieve an object with the highest rank with the complexity of \(\Theta(1)\), 
(2) insert a new object \(\Theta(1)\), and 
(3) delete an object with the complexity of \(\Theta(\log n)\). 
Furthermore, the use of a hash table will allow us to update any node \(\Theta(1)\).
	
As explained above, the number of messages per round is \(O(n)\) in the worst case. 
Every message may result in an update (delete and insert) of the respective node's rank in the $PL$ (\(\delta_{OnMsg}=O(\log n)\)). 
On a timeout event of $RoundTimer$, Algorithm~\ref{alg:le} may remove at most one entry from the $PL$ (\(\delta_{OnTimer}=O(\log n)\)). 
Assuming that the maximal message delivery time (\(\delta_d\)) is constant, according to Assumption~\ref{def:shortest-round} the worst case time complexity of a round is \(\delta_r=O(n\log n)\).

\section{Enhancements to the Leader Election Algorithm}
\label{sec:extensions}
In this section we provide a brief description of three enhancements of the proposed algorithm.
Application of the proposed algorithm in different environments may require some adjustments as described below.
	
\subsection{Merging Regions}
\label{sec:merge}
In contrast to the basic assumption of a single broadcast domain, dynamic changes in network topology may split and merge regions. 
Assume, for example, two separated LANs which have successfully executed the entire LE process. 
Merging between the two LANs may create a single broadcast domain (i.e., region) with two active leaders.
A similar situation may occur due to the violation of Assumption~\ref{ass:broadcast-reach}, or critical failures that may transfer the nodes to arbitrary invalid states.
	
In such a situation, the symmetry between the leaders must be broken, and one of them needs to be elected leader.
	
There are several possible parameters that can be used to break ties:
\begin{enumerate}
	\item Compare the number of nodes that performed a \emph{handshake} with each of the active leaders.
		
	\item Compare the number of rounds that each of these leaders was active in ($cntRounds$ parameter).
		
	\item Compare the accumulated ranks with which the leaders achieved their leadership (before the ranks were set to infinity in Line \ref{alg:le:infty}).
		
\end{enumerate}
In either case the comparison is easily made possible when including one of the abovementioned parameter in the leader's \emph{BeepMsg} and performing the rank comparison in lexicographic order.
	
We propose using the LE algorithm presented in this paper in order to choose the winning leader in cases involving multiple leaders.
	
\begin{conjecture}[Merging regions]
	\label{th:merge}
	Algorithm~\ref{alg:le} can be executed to break the symmetry among several leaders arising in the same region.
\end{conjecture}
	
After the winning leader is chosen, the losing leaders stop broadcasting $BeepMsg$ messages. 
As a result, their slave nodes perform a handshake with the winning leader.
	
\subsection{Self-Stabilization}
\label{sec:self-stab}
	
A distributed system that is self-stabilizing converges in a finite number of steps to a legitimate state no matter what arbitrary state it is initialized with \cite{dolev2000self,dijkstra1982self}. 
Moreover, a self-stabilizing system will remain in a legitimate state as long as there is no violation of this state.
	
Invalid states in the leader election process are those in which there are more than one leader or states in which a leader cannot be elected. 
We have shown that given Assumption~\ref{ass:stable-node}, at least one leader will be elected eventually.
In the previous sub-section we showed that merging is possible in cases in which there are two or more leaders. 
Thus, if our basic assumptions are no longer violated, Conjecture~\ref{th:merge} can be used to prove self-stabilization.
	
\begin{conjecture}[Self-stabilization]	
	\label{th:self*}
	Assuming that Conjecture~\ref{th:merge} is true, a leader election protocol which detects cases of multiple leaders (elected due to memory failures), and executes Algorithm~\ref{alg:le} to choose among them, ends up with a single leader within a finite time (following the last failure).	
\end{conjecture}
	
\subsection{Adapting to Wireless Ad Hoc Networks (WANETs)}
\label{sec:adhoc}
	
An ad hoc network is built spontaneously as nodes connect. 
Ad hoc networks do not rely on a physical infrastructure, such as routers in wired networks or access points in wireless networks.
	
In order for an ad hoc network to operate, each node is required to participate in the routing task by forwarding data to other nodes. 
The routing decisions are made dynamically on the basis of network connectivity.
	
There are two modifications to be made in order to adapt the proposed LE algorithm to ad hoc networks. 
Ad hoc networks rely on multi-hop connectivity, rather than a single hop (clique) and thus, our algorithm must be adjusted to accommodate multi-hop connectivity. 
First, each node participating in the LE process is required to forward every incoming message to its neighbors. 
This can be implemented in the network layer or as part of the \emph{BeepMsg received} handler method (see Algorithm~\ref{alg:le}, Line~\ref{alg:le:beepmsg}). 
Second, in an ad hoc network with diameter $D$, a messages maximal delivery time $\delta_d$ must be multiplied by $D$.
	
\section{Summary}
\label{sec:conc}
	
In this study we introduced the Partially Asynchronous Agile Leader Election algorithm (PALE) for distributed networks. 
The vast majority of state of the art algorithms for distributed leader election rely on strong assumptions regarding synchronization or network stability. 
In this paper, we obviate all the assumptions on network stability except the minimal requirement that there is at least one node that survives long enough to be a leader. 
We also significantly weaken the assumptions on the nodes' clocks; 
PALE only requires bounded clock drifts instead of the global clock, causal clocks, or distributed synchronization layer used in previous research. 
Overall, PALE achieves the best trade-off between synchronization and stability assumptions.
	
PALE operates well under harsh conditions such as desynchronized clocks and jittering nodes (that join or leave the network at arbitrary times), even when those nodes are highly ranked according to their physical parameters. 
The ability of PALE to handle jittering nodes is achieved by gradually increasing the ranks of active nodes. 
We prove that the algorithm satisfies uniqueness, agreement, and finite time termination requirements, even under the worst case conditions.
	
In the worst case scenario, PALE terminates after $O(n)$ rounds (according to the clock of the slowest node), while in every round a fraction of the nodes jitter and send a message. 
Under the monotonic behavior conditions, assumed by \citet{Larrea2011}, PALE requires $O(n)$ messages in total (amortized message complexity of $O(1)$ messages per round). 
Under mild conditions, when the strongest nodes don't leave or join the network, the substitution of a leader requires $O(1)$ messages in total. 
This message complexity is achieved due to the list of participants maintained by active nodes and the fact that only the forthcoming leader broadcasts its rank.
Implementation and showcase of PALE can be found in IEEE Code Ocean. 
	
We also described a few enhancements: 
(1) merging two or more regions with existing leaders, 
(2) supporting the self-stabilization property, and 
(3) adapting to wireless ad hoc networks (WANETs). 
In addition to the improvements outlined, future research directions may also include: nodes' mobility and a secured scheme that decreases the chances for a malicious node to manipulate its rank and become a leader.

\ifCLASSOPTIONcaptionsoff
  \newpage
\fi



\bibliographystyle{IEEEtranN}
\bibliography{IEEEabrv,leader-election}
\label{Bibliography}
\end{document}